\documentclass[journal]{IEEEtran}
\ifCLASSINFOpdf
\else
\fi
\usepackage{scalerel}
\usepackage{tikz}
\usetikzlibrary{svg.path}

\usepackage{graphicx}
\usepackage{amsfonts}
\usepackage{mathrsfs}
\usepackage{amsmath} 

\usepackage{scalerel}
\usepackage[ruled,vlined]{algorithm2e}

\usepackage{bm}
\usepackage{booktabs,ragged2e}

\usepackage[flushleft]{threeparttable}

\newtheorem{definition}{Definition}
\usepackage{tabularx}
\newcolumntype{C}{>{\centering\arraybackslash}X} 
\usepackage{threeparttable}
\usepackage{multirow}
\usepackage[utf8]{inputenc}
\usepackage{newunicodechar}
\usepackage{booktabs}

\usepackage{scalerel}
\usepackage{tikz}
\usetikzlibrary{svg.path}

\definecolor{orcidlogocol}{HTML}{A6CE39}
\tikzset{
  orcidlogo/.pic={
    \fill[orcidlogocol] svg{M256,128c0,70.7-57.3,128-128,128C57.3,256,0,198.7,0,128C0,57.3,57.3,0,128,0C198.7,0,256,57.3,256,128z};
    \fill[white] svg{M86.3,186.2H70.9V79.1h15.4v48.4V186.2z}
                 svg{M108.9,79.1h41.6c39.6,0,57,28.3,57,53.6c0,27.5-21.5,53.6-56.8,53.6h-41.8V79.1z M124.3,172.4h24.5c34.9,0,42.9-26.5,42.9-39.7c0-21.5-13.7-39.7-43.7-39.7h-23.7V172.4z}
                 svg{M88.7,56.8c0,5.5-4.5,10.1-10.1,10.1c-5.6,0-10.1-4.6-10.1-10.1c0-5.6,4.5-10.1,10.1-10.1C84.2,46.7,88.7,51.3,88.7,56.8z};
  }
}

\newcommand\orcidicon[1]{\href{https://orcid.org/#1}{\mbox{\scalerel*{
\begin{tikzpicture}[yscale=-1,transform shape]
\pic{orcidlogo};
\end{tikzpicture}
}{|}}}}

\usepackage{hyperref} 

\usepackage{graphicx}
\usepackage{lipsum}

\begin{document}
%
\title{Topological EEG Nonlinear Dynamics Analysis for Emotion Recognition}
%
%
%

\author{Yan~Yan$^{\orcidicon{0000-0002-6344-136X}}$, ~\IEEEmembership{Member,~IEEE,} Xuankun~Wu$^\dag$, Chengdong~Li, Yini~He, Zhicheng~Zhang, Huihui~Li, Ang~Li$^*$, and~Lei~Wang$^*$$^{\orcidicon{0000-0002-7033-9806}}$,~\IEEEmembership{Senior Member,~IEEE}
\thanks{ Y. Yan, X. Wu, C. Li, H. Li, and L. Wang are with the Shenzhen Institutes of Advanced Technology, Chinese Academy of Sciences, Shenzhen, Guangdong, 518055, China. E-mail: \{yan.yan, xk.wu, cd.li, hh.li, wang.lei\}@siat.ac.cn}
\thanks{Y. Yan, A. Li and L. Wang are also with University of the Chinese Academy of Sciences}
\thanks{Z. Zhang is with Department of Radiation Oncology, Stanford University, Stanford, CA, 94305, USA. E-mail: zzc623@stanford.edu}
\thanks{Y. He is with Beijing Normal University. E-mail: heyini1115@gmail.com}
\thanks{A. Li is with State Key Laboratory of Brain and Cognitive Science, Institute of Biophysics, Chinese Academy of Sciences. E-mail: al@ibp.ac.cn}
\thanks{$^\dag$ indicates joint first author. $^*$ indicates the corresponding authors. This work was supported in part by the xxx.}}

%
%

\markboth{~}%
{Yan {\textit{et al.}}: Topological EEG Nonlinear Dynamics Analysis for Emotion Recognition}
%




\maketitle

\begin{abstract}
Emotional recognition through exploring the electroencephalography (EEG) characteristics has been widely performed in recent studies.
Nonlinear analysis and feature extraction methods for understanding the complex dynamical phenomena are associated with the EEG patterns of different emotions.
The phase space reconstruction is a typical nonlinear technique to reveal the dynamics of the brain neural system.
Recently, the topological data analysis (TDA) scheme has been used to explore the properties of space, which provides a powerful tool to think over the phase space.
In this work, we proposed a topological EEG nonlinear dynamics analysis approach using the phase space reconstruction (PSR) technique to convert EEG time series into phase space, and the persistent homology tool explores the topological properties of the phase space. 
We perform the topological analysis of EEG signals in different rhythm bands to build emotion feature vectors, which shows high distinguishing ability. 
We evaluate the approach with two well-known benchmark datasets, the DEAP and DREAMER datasets. The recognition results achieved accuracies of 99.37\% and 99.35\% in arousal and valence classification tasks with DEAP, and 99.96\%, 99.93\%, and 99.95\% in arousal, valence, and dominance classifications tasks with DREAMER, respectively. The performances are supposed to be outperformed current state-of-art approaches in DREAMER (improved by 1\% to 10\% depends on temporal length), while comparable to other related works evaluated in DEAP. 
The proposed work is the first investigation in the emotion recognition oriented EEG topological feature analysis, which brought a novel insight into the brain neural system nonlinear dynamics analysis and feature extraction. 
\end{abstract}

\begin{IEEEkeywords}
EEG emotion recognition, affective computing, topological data analysis, nonlinear dynamics, phase space reconstruction, dynamical systems, biomedical signal processing.
\end{IEEEkeywords}

%
\IEEEpeerreviewmaketitle

\section{Introduction}\label{sec:introduction}

\IEEEPARstart{E}{motion} recognition plays a vital role in affective computing, which identifies the human emotional states from behavioral activities or physiological signals.
Accurately recognizing the human's emotional states significantly improve the reliability and intelligence level in human-computer interaction \cite{soleymani2011multimodal, knapp2011physiological}, healthcare monitoring \cite{ali2016eeg, choong2021hurst}, and behavior evaluation \cite{chai2019exploring} applications.
Physiological signals variation is spontaneous and complex to conceal when the emotional state changes, providing an ideal emotion recognition technique.
 EEG signals can be obtained easily with wearable systems measuring the voltage levels changes due to the ionic current flows variation in the neurons of the brain \cite{li2019robust}. 
As wearable technologies are dramatically developing, the EEG acquiring techniques provide a preferable way to explore brain responses to emotional stimuli. 
EEG-based emotion recognition has drawn an increasing amount of attention in recent years.

There are diverse emotion models proposed to describe emotional states. 
The discrete model categorized the emotional states into six discrete classes: anger, disgust, fear, happiness, sadness, and surprise.
The dimensional model considered emotions with arousal, valence, and dominance levels \cite{mehrabian1996pleasure}, which describe the degree from unpleasant to pleasant, passive to active, and submissive to dominant, respectively \cite{zhong2020eeg}.
In this work, we use the dimensional model in the emotion recognition tasks, namely the arousal, valence, and dominance levels (low/high), which formed the low/high arousal (LA/HA), low/high valence (LV/HV), and low/high dominance (LD/HD) categories.

EEG signals capture the brain activities with the electrodes placed at different head locations, which tracks the variations of different parts of encephalic regions.
The collected EEG signals are often investigated in the bands of $\delta$ (1-4 Hz), $\theta$ (4-8 Hz), $\alpha$ (8-13 Hz), $\beta$ (13-30 Hz), and $\gamma$ (greater than 30 Hz) \cite{zheng2017identifying, zheng2015investigating}. 
The EEG signals were first decomposed to the frequency bands, and then the feature extracting was performed.
Li et al. \cite{li2009emotion} proposed a gamma-band EEG-based emotions-happiness and sadness classification.
Shi et al. \cite{shi2013differential} introduced a differential entropy-based approach toward EEG-based vigilance estimating with the EEG bands.
Murugappan et al. \cite{murugappan2009appraising} considered the alpha-band EEG signal to build nonlinear features to classify the emotions.
This work considers the $\theta$, $\alpha$, $\beta$, and $\gamma$ band of EEG as used in most emotion recognition applications.

Generally, the emotional recognition tasks were performed with feature extraction and classification with different classifiers. 
With the frequency band EEG signals, the common used features are differential entropies \cite{shi2013differential, zheng2014eeg}, power spectral density \cite{frantzidis2010toward}, differential asymmetry parameters\cite{liu2013real}, the rational asymmetry features\cite{lin2010eeg} and the differential caudality \cite{zheng2015investigating}.
Meanwhile, the spatial and temporal features used to acquire temporal information in EEG-based emotion recognition, such as Hjorth feature \cite{hjorth1970eeg}, fractal dimension \cite{liu2013real}, higher order crossing feature \cite{petrantonakis2009emotion}, global field power temporal features\cite{jrad2012identification}, local-learning-based spatial-temporal components \cite{jiang2017spatial}, group sparse canonical correlation analysis \cite{zheng2016multichannel}, empirical mode decomposition \cite{mert2018emotion, liu2018electroencephalogram} and independent residual analysis \cite{zhao2007temporal} etc.
Recently, a variety of deep learning structures have been proposed to extract EEG features toward emotional recognition.
Zheng et al. \cite{zheng2015investigating} proposed a deep neural networks approach to investigate the critical frequency bands and channels for EEG-based emotion recognition.
Xin et al. \cite{chai2016unsupervised} combined an auto-encoder network and a subspace alignment solution in a unified framework toward EEG-based emotional state classification.
Cui et al. \cite{zhang2018spatial, cui2020eeg} introduced an end-to-end regional-asymmetric convolutional neural network.
Dynamical graph convolutional neural networks (DGCNN) \cite{song2018eeg}, and sparse DGCNN model which modifies DGCNN by imposing a sparseness constraint was introduced by \cite{zhang2021sparsedgcnn}. 
Zhong et al. \cite{zhong2020eeg} proposed a regularized graph neural networks-based method toward emotion recognition using EEG signals.
Recurrent models like reservoir computing \cite{fourati2020unsupervised}, attention-based convolutional recurrent neural network \cite{tao2020eeg} was also involved in the EEG-based affection computing.

Meanwhile, since EEG is generated by the brain system supposed to be highly complex, the acquired signals indicate nonlinearity, non-stationary and chaotic behavior \cite{soroush2020emotion}.
Nonlinear analysis of EEG signals has been widely performed and used to build features toward emotional recognition \cite{garcia2017nonlinear, soroush2018novel, fan2018recognizing, tong2017eeg, soroush2019emotion}.
Alcaraz et al. \cite{alcaraz2017recent} conclude the nonlinear characterization of EEG into the five following categories: (1) Fractal fluctuations quantifications, as proposed in \cite{liu2014eeg, paul2015eeg, hatamikia2014recognition}; (2) Irregularities quantifications by entropy parameters, such as the works proposed in \cite{molina2011comparative, jie2014emotion, garcia2016application, hosseini2010emotional, li2015improved}; (3) Information contents quantifications by using discrete symbols, typical examples are \cite{chen2017novel, li2016application}; (4) Chaos degree descriptors using PSR for feature extraction, such as Lyapunov exponents proposed in \cite{hoseingholizade2012studying, acar2015feature, natarajan2004nonlinear}; (5) Geometric representation of chaos developed in \cite{bahari2013eeg, yang2018recurrence, goshvarpour2016recurrence, soroush2020emotion}. The nonlinear characterization of EEG provided essential information of the brain state. The nonlinear descriptors widely adopted in EEG signal analysis show great discrimination ability in emotional states recognition.

The topological data analysis (TDA) scheme was recently proposed to represent point clouds' geometric structure, which inspired novel insights toward phase space information extraction.
The TDA technique adopts a persistent homology \cite{edelsbrunner2008persistent, otter2017roadmap} tool to describe the point clouds, providing a novel description of the structure of the point clouds and topological properties of the phase space.
The nonlinear dynamics analysis with topological descriptions has been used in wheeze detection \cite{emrani2014persistent}, heart dynamics analysis toward arrhythmia detection \cite{safarbali2019nonlinear}, gait dynamics analysis toward neurodegenerative disease discrimination \cite{yan2020gait, yan2020classification}, EEG-based dynamics analysis toward brain state recognition\cite{kilner2010topological, wang2018topological, piangerelli2018topological, wang2019statistical, altindics2021parameter, majumder2020detecting, wang2020topological, stolz2021topological} and plenty of time series classification applications \cite{seversky2016time, umeda2017time, khasawneh2018chatter}.
This work proposes a topological nonlinear dynamics analysis approach toward EEG-based emotion recognizing as a complement of the phase space information, namely topological EEG nonlinear dynamics analysis (TEEGNDA).
This work is supposed to be the first attempt at topological nonlinear analysis and topological machine learning in emotional state recognition and affective computing.
The main contributions of this work are as follows:
\begin{enumerate}
\item We proposed the topological nonlinear analysis of the multi-band EEG signals with the corresponding features to reveal the dynamical variation of different emotional states. The signals were first decomposed to the $\theta$, $\alpha$, $\beta$, $\gamma$ bands, and then the topological nonlinear analysis and feature extracting were performed separately. The topological features from each band were stacked to build vectors toward emotional state recognition.
\item We validated the single-channel EEG-based emotional recognition performance, which proved the recognition ability of the topological descriptors. 
The single-channel EEGs are used for sub rhythm band topological nonlinear dynamics analysis, achieving relative high recognition accuracies/standard deviations of 90.60/0.52 and 89.78/0.59 percents for arousal and valence classification in the DEAP dataset, while 98.51/0.36, 98.44/0.39, and 98.47/0.39 percents for arousal, valence, and dominance recognition in the DREAMER database.
\item We also illustrate the emotional recognition experiments including Low/High Valence discrimination (based on DEAP, DREAMER), Low/High Arousal discrimination (based on DEAP, DREAMER), and Low/High Dominance discrimination (based on DREAMER) using the channel-fusion strategy. 
The average accuracies of 99.37 and 99.35 percent are obtained for arousal and valence classification in the DEAP dataset, while 99.96, 99.93, and 99.95 percent are obtained for arousal and valence and dominance classifications on the DREAMER database. 
The results are supposed to comparative or outperform current models in the subject-wise experiments, which proved the distinguishing ability of the topological descriptions of phase space.
\end{enumerate}



\section{Preliminary of Topological Data Analysis}\label{sec:preliminary}

\subsection{Simplicial Complex}
\begin{figure}[]
\centering
\includegraphics[width=3.3in]{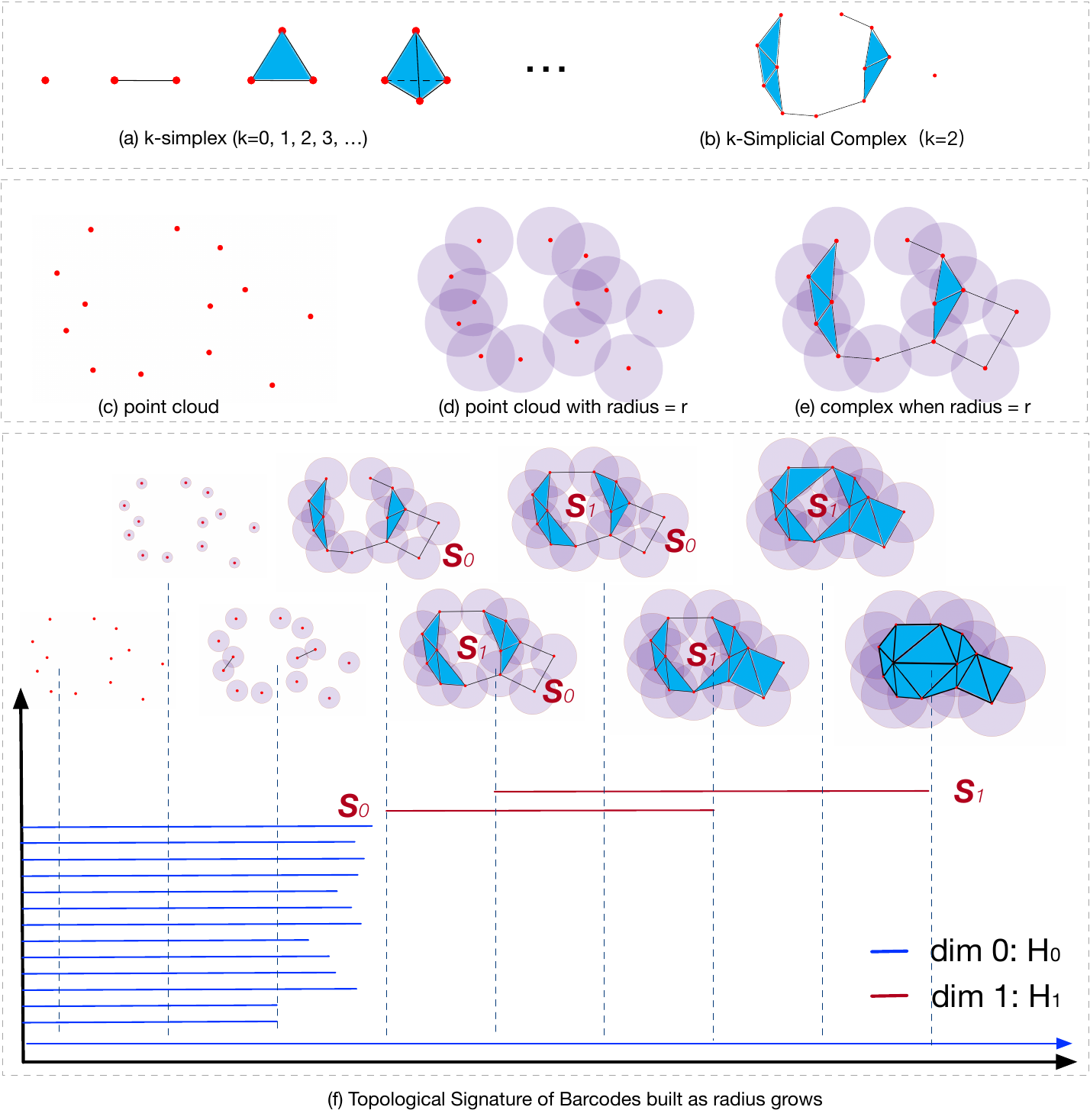}
\caption{Preliminary of topological data analysis: (a) the $k$-simplexes; (b) the combination of simplexes forms a simplicial complex; (c) a 2-dimensional point cloud used for illustration; (d) the points are turned into radius-based balls, namely $r$-ball; (e) simplicial complex used to describe the structure of $r$-balls; (f) barcodes achieved via gradually increasing the radius of $r$-balls, including the $0$-dimensional homology class i.e. $\mathbb{H}_0$ objects (connected components) illustrated in blue, and  $1$-dimensional homology class i.e. $\mathbb{H}_1$ objects (holes) illustrated in red (red bars correspond to $S_0$ and $S_1$).}
\label{fig_preliminary}
\end{figure}

Consider point set $\bm{X}$ in a space, any subset of a point cloud with cardinality $k+1$ is called a $k$-simplex \cite{rieck2017persistent}, as in the graph-theoretic context, the $0$-simplices are vertices, $1$-simplices are edges, $2$-simplices are triangular faces, and $3$-simplices are tetrahedrons (Figure \ref{fig_preliminary}.(a)).
One simplicial complexes (Figure \ref{fig_preliminary}.(b)) include all the lower dimensional simplices along with their highest dimension ones, thus one graph composed of vertices and edges is described as a $1$-dimensional simplicial complex.
Mathematically, 
\begin{definition}
A simplicial complex $\mathcal{R}$ is a finite collection of simplices, for each simplex $\sigma$,
\begin{enumerate}
\item any face of $\sigma \in \mathcal{R}$ is also in $\mathcal{R}$, and
\item if $\sigma_1, \sigma_2 \in \mathcal{R}$, then $\sigma_1 \cap \sigma_2$ is a face for both $\sigma_1$ and $\sigma_2$.
\end{enumerate}
\end{definition}
With the simplex and simplicial complex notations, point cloud $\bm{X}$ is converted into a simplicial complex with:
\begin{definition}
Given a scale parameter $\epsilon$ and a point cloud $\bm{X}$, the Vietoris-Rips complex $\mathcal{R}(\bm{X}, \epsilon)$ is defined as a simplicial complex contains all subsets with maximum diameter $\epsilon$:
\begin{equation}
\mathcal{V}(\bm{\epsilon}):=\{\sigma \subseteq \bm{X} | \mbox{diam} \sigma \leq r\}
\end{equation}
\end{definition}
in which $\mbox{diam} \sigma$ means:
\begin{definition}
Let $\mathbb{T}$ be a finite topological space with a metric of $dist_{\mathbb{T}}$. The diameter of  $\mathbb{T}$ is the upper bound of the set of all pairwise distances, i.e.
\begin{equation}
\mbox{diam} \mathbb{T} := \mbox{sup} \{\mbox{dist}_\mathbb{T}(x, y) |x, y \in \mathbb{T}\}
\end{equation}
\end{definition}
Since the topological space $\mathbb{T}$ is finite, the supremum is attained by some pairs of point, and alway exists when we investigate a point cloud with a finite number of point.
Thus, the complex $\mathcal{R}$ contains a simplex $\sigma = \{v_0, v_1, \ldots\}$ if and only all the points $v_0, v_1, \ldots$ are within a distance of at most $r$ to each other.

The scale parameter $\epsilon$ is a variable ranging from $0$ to $\infty$.
With the Vietoris-Rips notation, the origin point cloud is $\mathcal{R}(\bm{X}, 0)$, while all points merge in $\mathcal{R}(\bm{X}, \infty)$.
The topological properties of the space which the points lying on can be characterized via tracking the  Vietoris-Rips complex while gradually increasing the scale parameter, namely the persistent homology technique.

\subsection{Persistent Homology}
Consider the point cloud $\mathbf{X}$ with $m$ points $\{\mathbf{x}_1, \mathbf{x}_2, \ldots \mathbf{x}_m\}$.
\begin{enumerate}
\item First, replace the points of $\{\mathbf{x}_1, \mathbf{x}_2, \ldots \mathbf{x}_m\}$ with the radius-based sphere (circle in 2-D case, as in Figure \ref{fig_preliminary}.(d)) $\{B(\mathbf{x}_1, \epsilon), B(\mathbf{x}_2, \epsilon),  \cdots, B(\mathbf{x}_m, \epsilon)\}$.
\item Then, gradually increase the radius  $\epsilon$ from $0$ to $\infty$. As $\epsilon$ increasing, the $\epsilon$-spheres may merge to form new components, and holes (Figure \ref{fig_preliminary}.(e)).
\item Finally, all $B(\bm{X}, \epsilon)$ object merge into one component when $\epsilon$ value turns large enough.
\end{enumerate}

The components and holes appear and disappear as illustrated in Figure \ref{fig_preliminary}.(f), the connected components belong to the $0$-dimensional homology class $\mathbb{H}_0$, while the holes belong to the 1-dimensional homology class $\mathbb{H}_1$.
The $\mathbb{H}_1$ instances are $\bm{S}_0$ and $\bm{S}_1$, which denotes the hole located at the middle and bottom right of the complex, respectively.

The growing process with increasing radius parameters $\{\epsilon_0, \epsilon_1, \epsilon_2, \ldots \} \in \bm{\epsilon}$:
\begin{equation}
\mathcal{B}(\bm{X}, \epsilon_0), \mathcal{B}(\bm{X}, \epsilon_1), \mathcal{B}(\bm{X}, \epsilon_2), \mathcal{B}(\bm{X}, \epsilon_3), \ldots
\end{equation}
is represented as a sequence of complexes:
\begin{equation}
\mathcal{R}(\bm{X}, \epsilon_0), \mathcal{R}(\bm{X}, \epsilon_1), \mathcal{R}(\bm{X}, \epsilon_2), \mathcal{R}(\bm{X}, \epsilon_3), \ldots
\end{equation}
Meanwhile, the subsequent Rips complex in the sequence is larger than its previous ones, which is  as \textit{nested}. 
The nested Rips complex sequence is called a \textit{filtration}, which has the property that
\begin{equation}\label{equ:filtration}
\mathcal{R}(\bm{X}, \epsilon_0) \subseteq \mathcal{R}(\bm{X}, \epsilon_1) \subseteq \mathcal{R}(\bm{X}, \epsilon_2) \subseteq \ldots \subseteq \mathcal{R}(\bm{X}, \epsilon_n)
\end{equation}
when $\epsilon_0 \leq \epsilon_1 \leq \ldots \epsilon_n$.
Thus, for each point cloud embedded from the time series, we have a Vietoris-Rips complex sequence with the varying $\epsilon$, i.e., Vietoris-Rips filtration (the theoretical introduction and implementation algorithm of building Vietoris-Rips complex from point cloud are described detailedly in \cite{zomorodian2010fast}).
Through tracking the growing process, the birth-death ordered pairs for the homology objects are recorded as persistence of the homology. Mathematically, 
\begin{definition}
Let $\mathbb{H}$ be a homology class that get created in $\mathcal{R}(\bm{X}, \epsilon_i)$ and destroyed in $\mathcal{R}(\bm{X}, \epsilon_j)$, the corresponding filtration values are $\epsilon_i, \epsilon_j$. Then we say the homology class $\mathbb{H}$ has a \textit{persistence} of:
\begin{equation}
\mbox{pers}(\mathbb{H}):=\epsilon_j - \epsilon_i
\end{equation}
\end{definition} 

The persistences and birth-death ordered pairs can be visualized using \textit{barcodes}, which track the filtration values of the birth time and death time for each homology object in the nested sequence.
The blue bars in the barcodes plot denote the persistence of connected components $\mathbb{H}_0$, while the red bars represent holes $\mathbb{H}_1$ (Figure \ref{fig_preliminary}.(f)). 
For higher dimensional phase space, the dimension of the homology class increase accordingly. 
However, in this work we focus only on three low dimensional homologies, i.e., $\mathbb{H}_0$, $\mathbb{H}_1$, and $\mathbb{H}_2$.

\subsection{Topological Summaries}
The persistence homology technique extracts the topological characteristics of the point cloud via recording the lifetime of the objects in different homology class.
Consider the homology classes with dimensions of $0$, $1$, and $2$, namely $\mathbb{H}_0$, $\mathbb{H}_1$, and $\mathbb{H}_2$.
The homology numbers are $A$, $B$, $C$ for $\mathbb{H}_0$, $\mathbb{H}_1$, and $\mathbb{H}_2$, respectively.
Thus we have three sets to represent the topological features, for $\mathbb{H}_0$ the topological summaries are:
\begin{equation}
\{\{b_0^1, d_0^1\}, \{b_0^2, d_0^2\}, \cdots \{b_0^A, d_0^A\}\}
\label{betti_0}
\end{equation} 
while for $\mathbb{H}_1$ we have
\begin{equation}
\{\{b_1^1, d_1^1\}, \{b_1^2, d_1^2\}, \cdots \{b_1^B, d_1^B\}\}
\label{betti_1}
\end{equation} 
and for $\mathbb{H}_2$ the summaries are represented with
\begin{equation}
\{\{b_2^1, d_2^1\}, \{b_2^2, d_2^2\}, \cdots \{b_2^C, d_2^C\}\}
\label{betti_2}
\end{equation}
Figure \ref{fig_preliminary}.(f) illustrates the barcodes demonstrations of $\mathbb{H}_0$ and $\mathbb{H}_1$.  
A variety of parameters or feature sets were proposed in previous studies, such as the statistical properties, distance analysis, rule-based features, kernels built based on the topological summaries.
In this work, we consider the persistence landscapes (PLs) as extracted topological features based on the $\mathbb{H}_0$, $\mathbb{H}_1$, and $\mathbb{H}_2$ of the point cloud, we put the details of PL in Section \ref{sec:methodology}.

\section{TEEGNDA for Emotion Recognition}\label{sec:methodology}
\subsection{Framework of Proposed TEEGNDA}

\begin{algorithm}[]\label{alg_teegnda}
\SetAlgoLined
\textbf{Input:} $N$ preprocessed training samples with $M$-channel based four sub-bands EEG signal segments $\mathbf{T}^{N \times M \times 4} $, with labels of $\hat{\mathbf{y}}$, PSR dimension of $d$ and time lag of $\tau$, point cloud topological analysis involved homology classes $H = 0, 1, 2$\\
\textbf{Output:} EEG-based emotion recognition feature set and label set $\{\mathbf{F}, \hat{\mathbf{y}}\}$  \\
1:\textbf{for} $i = 1:T$  \\
2:$ \quad $ \textbf{for} $j = 1:M$  \\
3:$\quad   \quad $ \textbf{for} $l = 1:4$ \\
4:$ \quad \quad \quad $ (1) Embed time series $\bm{t} = \{t_1, t_2, \ldots, t_w\}$ with $d$ and $\tau$ to one point cloud $\mathbf{X}$ as in Equation \ref{equ:time_delay}  \\
5:$\quad \quad \quad $ (2)  Perform the persistent homology process as to build filtration as in \ref{equ:filtration} \\
6:$\quad \quad \quad $ (3)  Record the persistences and barcodes of $\mathbb{H}_0$, $\mathbb{H}_1$, and $\mathbb{H}_2$ as in \ref{betti_0} and \ref{betti_1}   \\
7:$\quad \quad \quad $ (4)  Extract the PL of the $\mathbb{H}_0$, $\mathbb{H}_1$, and $\mathbb{H}_2$ objects from the barcodes with Equation \ref{equ:pl_function} and  Equation \ref{equ:pl_function_2} \\
7:$\quad \quad \quad $ (5)  Compute the average values of the PLs for each homology class as $\mathbf{F}_{ijl}$\\
8:$\quad \quad  $ \textbf{return} $\mathbf{F}_{ijl}$ \\
9:$\quad  $ \textbf{return} $\mathbf{F}_{ij} = \{\mathbf{F}_{ij\theta}, \mathbf{F}_{ij\alpha}, \mathbf{F}_{ij\beta}, \mathbf{F}_{ij\gamma}\}$ \\
10:\textbf{return} $\mathbf{F}_{i}$
\caption{Feature Extratcion in EEG-based Emotion Recognition using TEEGNDA}
\end{algorithm}

\begin{figure*}[]
\centering
\includegraphics[width=7.2in]{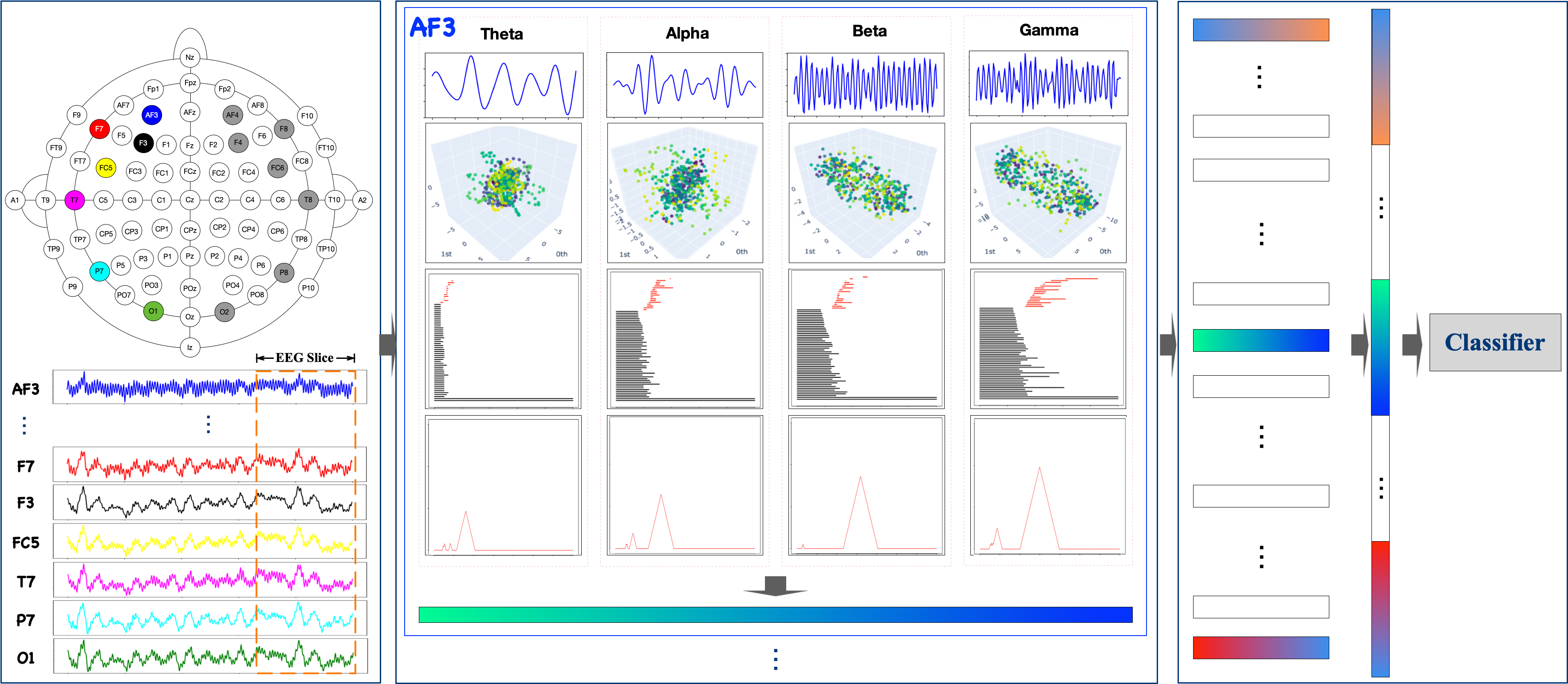}
\caption{The framework of the TEEGNDA model for EEG emotion recognition consists of the phase space reconstruction via time-delay embedding, Barcode extraction with the persistent homology modeling, topological feature generation with barcodes, and RF classification. The inputs of the model are the EEG segments from each channel. Each channel contains four frequency bands ($\theta$, $\alpha$, $\beta$, and $\gamma$). The outputs are the predicted labels through the RF classifier.}
\label{fig_system_frame}
\end{figure*}

The EEG-based emotion recognition approaches include signal pre-processing, signal segmenting, feature extracting, and then used as input of the classifier to classify the emotional states.
In this work, the pre-processed and segmented EEG data are firstly used to build feature sets and then recognized with the classifier.
We extract the topological features from the EEG with three stages: rhythm band extraction, phase space reconstruction (PSR), topological summaries-based feature extraction, and feature-based emotion classification.
The EEG signals are firstly denoised, and then four rhythm bands of $\theta$, $\alpha$, $\beta$, and $\gamma$ are extracted.
We use the PSR technique to convert the time series into the phase space via time-delay embedding for each EEG signal slice.
Then we have 4 point clouds revealing the nonlinear dynamics from four different frequency bands.
With the persistent homology technique introduced in Section \ref{sec:preliminary}, we extract the topological summaries from the point clouds.
Separately, we build the persistence landscape features based on the topological information of $\mathbb{H}_0$, $\mathbb{H}_1$, and $\mathbb{H}_2$ from the point clouds.
The rhythm band-based persistence landscapes are stacked to build the topological features for each EEG channel.
Finally, the band-based topological features are stacked to build the feature vectors and then used in the random forests classifier-based recognition system, with which the emotional recognition model is built and evaluated.
We term the whole process as topological EEG nonlinear dynamics analysis (TEEGNDA) toward EEG-based emotion recognition.
An overview of the proposed TEEGNDA approach is illustrated in Figure \ref{fig_system_frame}.
Meanwhile, we illustrate the TEEGNDA algorithm in Algorithm \ref{alg_teegnda}.

\subsection{Pre-processing and Sub-band Extraction}
The EEG signals include low and high-frequency noise, which is useless in the emotion recognition task. Thus we consider a band-pass filter with cut-off frequencies of $1$Hz and $75$Hz.
The pre-processed EEG signals contain the information within frequency from $1$Hz to $75$Hz.
We extract the four frequency bands of $\theta$, $\alpha$, $\beta$, and $\gamma$, with band-pass filters with cut-off frequencies of $1$-$4$Hz, $4$-$7$Hz, $8$-$13$Hz, and $13$-$30$Hz, respectively.
Then the sub-band rhythm-based EEG signals are segmented with the sliding windows.
There are four sub-band signals extracted for each channel to perform further PSR and corresponding topological feature extraction described in the following part.

\subsection{Phase Space Reconstruction}
A standard strategy for PSR is delay-coordinate embedding, with which the time series from a dynamical system is used to form vector-based points in phase space.
Mathematically, for time series $\mathbf{t} = \{t_1, t_2 \ldots t_w\}$, the delay-coordinate embedding process is denoted as
\begin{equation}\label{equ:time_delay}
\mathbf{x}_k = (t_k, t_{k + \tau}, \ldots, t_{k + (d-1)\tau})
\end{equation} 
in which $\tau$ means the time delay parameter, while $d$ is the dimension of the phase space, $k$ denotes the point index in the point cloud.
While the real-world sensors are time-length limited and interfered with by measurement noise, suitable delay-coordinate embedding parameters of $\tau$ and $d$ are needed to unfold the dynamics.
We suggest \cite{bradley2015nonlinear} for the discussions of the PSR in nonlinear time series analysis.
We use the average mutual information approach (AMI) \cite{fraser1986independent} toward choosing optimal $\tau$, while the false near neighbor (FNN) algorithm \cite{kennel1992determining} for $d$ selection.
Based on the recognition results of preliminary experiments, we choose the fixed parameters of $\tau = 8$, while $d = 10$.

\subsection{Topological Features Extraction}
\begin{figure}[]
\centering
\includegraphics[width=3.2in]{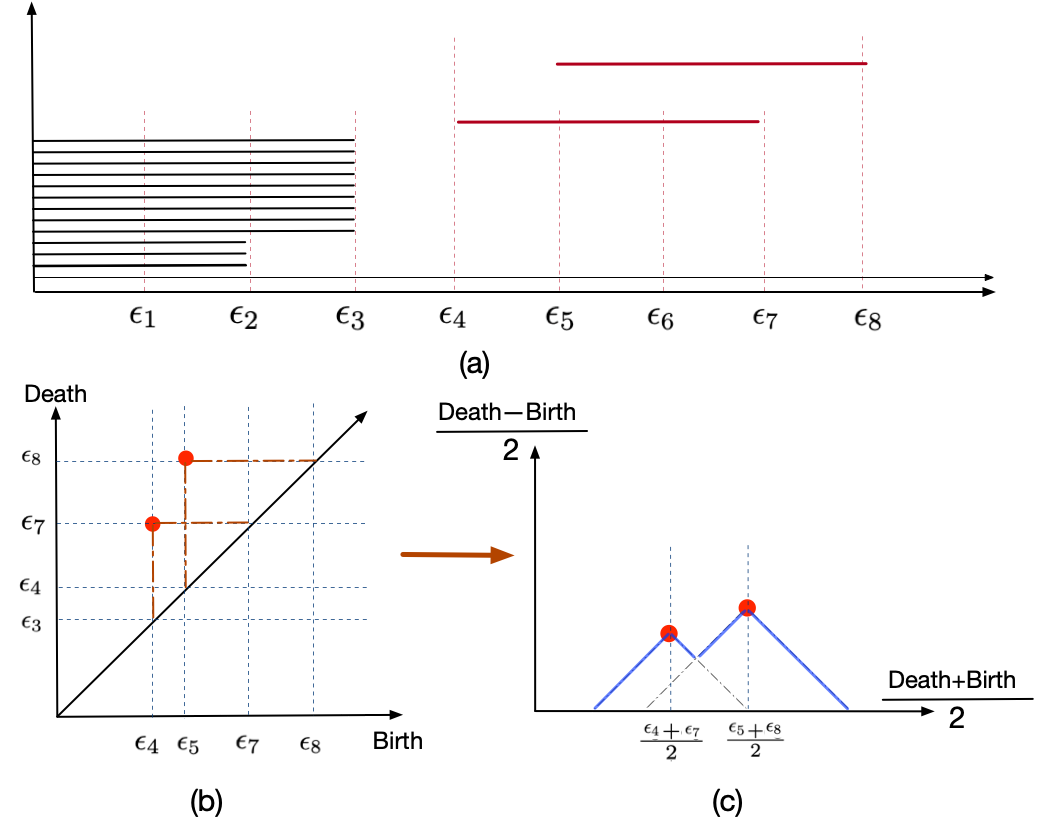}
\caption{Persistence landscapes developed from the barcodes: (a) barcodes examples of $\mathbb{H}_0$ (dark bars) and $\mathbb{H}_1$ (red bars); (b) persistence diagrams for the two $\mathbb{H}_1$ barcodes; (c) persistence landscapes of the red bars.}
\label{fig_pl}
\end{figure}

 The point cloud generated with the PSR technique reveals the dynamics of the nonlinear system. 
 As described in Section \ref{sec:preliminary}, the persistent homology tools develop topological descriptors of the nonlinear dynamics from the point cloud in the phase space.
In this work, we consider the lower dimensional homology classes of $\mathbb{H}_0$, $\mathbb{H}_1$, and $\mathbb{H}_2$, the corresponding instances of the topological summaries are illustrated in Equation \ref{betti_0}, \ref{betti_1}, and \ref{betti_2}, respectively.
An instance of barcodes illustrated \ref{betti_0} and \ref{betti_1} are  illustrated in Figure \ref{fig_preliminary}.(f) and \ref{fig_pl}.(a). 
The barcodes plots are further converted into persistence diagrams, which illustrates the persistence of homology object as point (horizontal axis as birth, while vertical for death) (Figure \ref{fig_pl}).
In this work, we use PLs extracted from the sub-band point clouds.
The main technical advantage of PL descriptor is that it is piecewise-linear functions that form feature vectors that are faster than using corresponding calculations with barcodes or persistence diagrams \cite{bubenik2015statistical}.

Mathematically, consider the point
\begin{equation}
 p_{pl} = (x, y) = \left( \frac{b+d}{2}, \frac{d-b}{2} \right)
\end{equation}
in which, $b$ for birth time while $d$ for death time.
We tent each point with the function 
\begin{equation}\label{equ:pl_function}
\bm{\Lambda}_p(t) = \left\{
\begin{array}{ll}
 t - x + y = t - b & t \in [x-y, x] \\
 x + y - t = d - t & t \in (x, x+y)\\
 0 & \mbox{otherwise}
\end{array}\right.
\end{equation}

Formally, PL of a persistence diagram $\mathcal{D}$ is a collection of functions as:
\begin{equation}\label{equ:pl_function_2}
\lambda_{\mathcal{D}}(\mbox{\textit{k}}, t) =\mbox{\textit{k}-max}_{p \in \mathcal{D}} \bm{\Lambda}_p(t), \quad t \in [0, T], \mbox{\textit{k}} \in \mathbb{Z}^+
\end{equation}
where \textit{k}-max is the \textit{k}-th largest value in the set, in this work we use $\mbox{\textit{k}}=1$ for the maximal value.

For an intuitive understanding of the PLs, we consider the two $\mathbb{H}_1$ objects' barcodes information represented as two red bars in Figure \ref{fig_pl}.(a), namely $\{\{\epsilon_4, \epsilon_7\}, \{\epsilon_5, \epsilon_8\}\}$.
The barcodes plot is converted into persistence diagrams as Figure \ref{fig_pl}.(b), which uses the birth parameters as the horizontal axis, while that the endpoint of the barcodes as the vertical axis.
Thus, the barcodes are turned into points $(\epsilon_4, \epsilon_7)$ and $(\epsilon_5, \epsilon_8)$ in Figure \ref{fig_pl}.(b).
Finally, the PLs are achieved via a rotate of the diagonal and the cumulative for the corresponding dimension of homologies, such as the two $\mathbb{H}_1$ objects in Figure \ref{fig_pl}.(c) with the blue silhouette curve.
The advantage of the persistence landscape representation is that the barcodes and persistence diagrams are mapped as elements of functional space to make it possible to perform the statistical analysis and build machine learning models. Other theoretical analysis and advantages discussions can be referred to in \cite{chazal2014stochastic}. 
In this work, we use the average value of PLs of $\mathbb{H}_0$, $\mathbb{H}_1$, and $\mathbb{H}_2$ as our topological features, which are used as the input for the classifier.

\subsection{Classification with Topological Features}
In this work, we consider the following experiments to illustrate the distinguishing ability of the proposed topological approach:
\begin{enumerate}
\item Exp. \#1: TEEGNDA uses all available channel fusion strategies with four frequency bands EEG to perform emotional states. We use several popular classifiers combined with the extracted emotion feature vectors, including the Gaussian Naive Bayes (GNB) classifier, K-nearest neighbor (kNN) classifier, Logistic Regression (LR) classifier, support vector machine (SVM) classifier, and Random Forests (RF) classifier.
\item Exp. \#2: TEEGNDA using single frequency band EEG of all available channel for emotion recognition to compare the rhythm band discrepancies.
\item Exp. \#3: TEEGNDA with different sliding window sizes towards validations and comparison with other related works.
\item Exp. \#4: TEEGNDA using single-channel EEG with four frequency bands to compare the channel differences, and evaluation the model effectiveness in single-channel occasions.
\item Exp. \#5: TEEGNDA evaluations with multiple emotion class recognition, which includes a 4-class classification in DEAP ( LALV, LAHV, HALV, and HAHV) with a threshold of 5; and an 8-class classification in DREAMER (HVLALD, HVLAHD , HVHALD, HVHAHD, LVLALD, LVLAHD, LVHALD, and LVHAHD, which denotes emotion states of protected, satisfied, surprised, happy, sad, unconcerned, frightened, and angry, respectively \cite{fourati2020unsupervised}) with a threshold of 3.
\end{enumerate}
The details of the experimental implementations and results are presented in the following sections.



\section{Experiments}\label{sec:experiments}
To validate the proposed approach, we conduct the experiments on two widely used databases, including the DREAMER database and DEAP database, both include multiple channels of EEG recordings. 
First, we introduce the involved datasets; then we demonstrate the model implementations; finally, we present the experiments and results of Exp. \#1, \#2, \#3, \#4, and \#5, respectively.

\subsection{Data Materials}
The DEAP dataset includes physiological signals of 32 subjects (16 males and 16 females), which are recorded when watching 40 music videos.
There are 32-channel EEG signals and other 8-channel physiological signals, in which only the EEG data are involved in the experiments.
The EEG rate is resampled from 512Hz to 128Hz, and the electrooculography artifacts were removed using the blind source separation technique.
The 40 one-minute clips were used to affect the participant's emotional state, with the self-assessment levels of arousal, valence, liking, and dominance for each video from 1 to 9 recorded.
Details of the DEAP dataset can be referred to in \cite{koelstra2011deap}.
We select the valence and arousal classification tasks as our model assessment criteria with a threshold value of 5 (LV when valence score less than 5, and HV when greater than 5, the similar setting for LA/HA).
Thus we have two binary classification tasks for the DEAP dataset, and we use DEAP-V and DEAP-A as the abbreviations of valence classification and arousal classification tasks, respectively.

The DREAMER database is a multimodal database including EEG and ECG recordings when the subjects were audio-visual stimulated.
Twenty-three subjects (14 males and 9 females) were asked to record the self-assessment levels (1 to 5) of arousal, valence, and dominance after each stimulus.
The EEG signals were recorded with a sampling frequency of 128Hz, while most of the artifacts were removed with linear phase FIR filters.
The involved film clips' lengths ranging from 65 seconds to 393 seconds, which are used to arousal emotional states, the total number of the video used is 18.
The locations of the headset are aligned according to the International 10-20 system: AF3, F7, F3, FC5, T7, P7, O1, O2, P8, T8, FC6, F4, F8, AF4, M1, and M2 \cite{badcock2013validation}. 
The mastoid sensor at M1 acted as a ground reference point for comparing the voltage of all other sensors, while the mastoid sensor at M2 was a feed-forward reference for reducing external electrical interference. 
Details of the DREAMER dataset can be referred to in \cite{kassa2013electrical}.
Thus, the signal from the other 14 contact sensors was recorded and used for feature extraction.
We choose the valence, arousal, and dominance levels to evaluate the models with a threshold value of 3.
Similarly, we use DREAMER-V, DREAMER-A, and DREAMER-D as the abbreviations for the three tasks in DREAMER, respectively.

\subsection{Implementations}
In this work, we only use the EEG signals from both datasets.
After the preprocessing stage, in the DEAP dataset, we have 40 $1$-minute long time series for each subject, while in DREAMER, we have 18 time series ($65s$ to $393s$ long) for each subject.
Each time series is segmented with specific overlap settings using fixed sample length (details in the description of the following experiments).
We shuffle all the segmented samples from different trials for each subject to build the training/testing sets, 80\% used for training and 20\% used for testing.
Then, we use 10-fold cross-validation to assess the performance of the proposed model. The mean classification accuracies with standard deviations based on subject-specified experiments are used as our model assessment criterion.
We use the Python package of \textit{giotto-tda} \cite{tauzin2021giotto} to perform the topological feature extraction, and \textit{scikit-learn} \cite{pedregosa2011scikit} in classification and cross-validation. 
Most of the classifier parameters are based on the default parameters in the packages without further tuning.


\begin{table*}[]
\caption{Exp \#1: Model Evaluation with Different Classifiers (with 1s window \& all bands).}
\begin{center}
\begin{tabular}{llccccccc}
\toprule
Dataset &  GNB(\%)	&   kNN(\%)	&  LR(\%)	 &  RF(\%)	 &  SVM(\%)	  \\
\midrule
DEAP-A & 61.62/5.67 & 85.73/5.82	& 90.44/3.56 & \textbf{99.37/0.73} &90.64/3.56   \\
DEAP-V & 60.14/5.12 & 83.95/6.12	& 89.65/3.47 & \textbf{99.35/0.91} & 89.85/3.47  \\
\midrule
DREAMER-A & 76.68/8.17 & 97.86/1.35 &  98.10/1.08 &  \textbf{99.96/0.07} & 98.52/1.08  \\
DREAMER-V &  75.96/6.95 & 97.89/1.37 & 98.05/1.22 & \textbf{99.93/0.07} & 98.49/1.22   \\
DREAMER-D & 74.80/7.14 & 97.99/1.17	& 98.04/1.02 & \textbf{99.95/0.07} & 98.38/1.02  \\
\bottomrule
\end{tabular}
\end{center}
\label{table_classifier}
\end{table*}


\begin{table*}[]
\caption{Exp \#2: Performance Comparison Using Different Rhythm Band Settings.}
\begin{center}
\begin{tabular}{ccccccc}
\toprule
Dataset\&Recognition Tasks & $\theta$-Band (\%) & $\alpha$-Band (\%) & $\beta$-Band (\%) &$\gamma$-Band (\%) & All Bands ($\theta$, $\alpha$, $\beta$, $\gamma$) (\%) \\
\midrule
DEAP-A & 97.27/2.77	&  99.13/0.86  & 83.29/5.37  &  86.38/5.19 & \textbf{99.37/0.73} \\
DEAP-V  & 97.37/2.83	& 99.20/0.95  & 82.08/5.88  & 85.72/5.57 &  \textbf{99.35/0.91} \\
\midrule
DREAMER-A & 99.81/0.34	& 99.67/0.77  &98.16/1.46    &97.42/2.16 & \textbf{99.96/0.07} \\
DREAMER-V & 99.86/0.36	& 99.72/0.67  &98.22/1.46  & 97.61/1.76 & \textbf{99.93/0.07}   \\
DREAMER-D & 99.83/0.51	& 99.81/0.46 & 98.08/1.43  & 97.45/1.68  & \textbf{99.95/0.07} \\
\bottomrule
\end{tabular}
\end{center}
\begin{tablenotes}
\RaggedRight
\item[a]  
\end{tablenotes}
\label{table_rhythm}
\end{table*}

\begin{table*}[]
\caption{Exp \#3: Number of Experiment Samples with Different Windows.}
\begin{center}
\begin{tabular}{llccccccc}
\toprule
Dataset &  0.5s (\%)	&  1s (\%)	& 2s (\%) & 3s (\%) & 4s (\%)  \\
\midrule
DEAP-A & 97.93/0.03 &\textbf{99.37/0.73}	& 95.17/0.45    & 89.03/4.83 &74.14/0.60   \\
DEAP-V &  98.28/0.03 &\textbf{99.35/0.91} & 94.52/0.73    & 89.04/4.53 &75.56/0.72 \\
\midrule
DREAMER-A & 99.93/0.09 &\textbf{99.96/0.07}	& 99.53/0.06  &98.70/0.20&98.32/0.19\\
DREAMER-V & 99.92/0.10 &\textbf{99.93/0.07} 	& 99.41/0.08  &97.35/0.30&97.16/0.25  \\
DREAMER-D & 99.82/0.53 &\textbf{99.95/0.07}    & 99.65/0.08 & 98.83/0.18&98.87/0.30  \\
\bottomrule
\end{tabular}
\end{center}
\label{table_win}
\end{table*}

\subsection{Exp. \#1: Emotion Recognition with TEEGNDA}
In Exp. \#1, four rhythm bands from the preprocessed signal, including $\theta$-band, $\alpha$-band, $\beta$-band, and $\gamma$-band, are extracted. We perform the PSR and topological feature extraction from each rhythm separately to build the feature vector.
The same procedures are performed to extract the topological features from the 32 channels of EEG in DEAP and 14 channels of EEG in DREAMER.
For each band signal, we use $d=8$ and $\tau = 10$ as our PSR parameters to convert the signals into point clouds.
Thus, the PLs are extracted from the point clouds.
We set the PL distribution ranges as 50 for each sub-band frequency signal and point cloud, and then we have a $200$-D feature vector for each channel, which means the dimensions of the final feature vectors are $6400 = 200 \time 32$ and $2800 = 200 \times 14$ for DEAP and DREAMER, respectively.
At the same time, we use the $1s$ temporal window size (namely 128 points since the sampling frequency is 128Hz) with a 25\% overlap to perform the EEG signal segmentation. 
For each subject, we use the TEEGNDA approach to distinguish the emotional states of the arousal and valence in DEAP, and arousal, valence and dominance in DREAMER as previous work in \cite{tao2020eeg}.

As illustrated in Table \ref{table_classifier}, the achieved best average accuracy/standard deviations(\%) for 32 subjects in DEAP and 23 subjects in DREAMER are 99.37/0.73, 99.35/0.91, 99.96/0.07, 99.93/0.07, and 99.95/0.07 for DEAP-A, DEAP-V, DREAMER-A, DREAMER-V, and DREAMER-D task, respectively(details for each subject are shown in Table A.I, A.II and Table A.III.). The best results are based on the RF classifier. Thus we only consider the RF classifier in the following experiments.




\subsection{Exp. \#2: Emotion Recognition with TEEGNDA Based on Single Rhythm Band}
In previous emotion recognition models, extracting information from the signals from different rhythm bands of EEG provides meaningful features to distinguish the emotional states.
Thus, we consider comparing the classification ability with different EEG rhythm bands (here $\theta$, $\alpha$, $\beta$, and $\gamma$ bands are involved).
As shown in Table \ref{table_rhythm}, the results of emotion recognition tasks using different rhythm bands are illustrated.
For the DEAP dataset, the emotion recognition results for LA/HA are 97.27/2.77(\%), 99.13/0.86(\%), 83.29/5.37(\%), and 86.38/5.19(\%) with $\theta$-band, $\alpha$-band, $\beta$-band, and $\gamma$-band, respectively.
The best performance accomplished is based on combining the four bands, namely the Total-column with 99.37/0.73(\%), which is better than the single rhythm solution.
The emotion recognition results for LA/HA classification in DEAP are 97.37/2.83(\%), 99.20/0.95(\%), 82.08/5.88(\%), and 85.72/5.57(\%) with $\theta$-band, $\alpha$-band, $\beta$-band, and $\gamma$-band, respectively.
We can see that the fusion of the four rhythm bands is better performed than the single ones, as shown in the Total-column with an accuracy/standard deviation(\%) of 99.35/0.91.

Meanwhile, for the DREAMER dataset, the emotion recognition task of LA/HA in DREAMER are 99.81/0.34(\%), 99.67/0.77(\%), 98.16/1.46(\%), and 97.42/2.16(\%) with $\theta$-band, $\alpha$-band, $\beta$-band, and $\gamma$-band, respectively.
The best performance was accomplished in the Total-column with an accuracy/standard deviation(\%) of 99.96/0.07.
The emotion recognition task of DREAMER-V is 99.86/0.36, 99.72/0.67, 98.22/1.46, and 97.61/1.76 with $\theta$-band, $\alpha$-band, $\beta$-band, and $\gamma$-band, respectively.
The best performance was accomplished in the Total-column with an accuracy/standard deviation(\%) of 99.93/0.07.
The results of the emotion recognition task of DREAMER-D are 99.83/0.51, 99.81/0.46, 98.08/1.43, and 97.45/1.68 with $\theta$-band, $\alpha$-band, $\beta$-band, and $\gamma$-band, respectively.
The best performance was accomplished in the Total-column with an accuracy/standard deviation(\%) of 99.95/0.07.

\subsection{Exp. \#3: Emotion Recognition with TEEGNDA with Different Sliding Window Size}
In Exp \#3, we consider the model's performance with three kinds of temporal windows with different lengths (i.e., $1s$, $2s$, and $4s$), the overlap for segmentation is $25\%$, and $3s$ with the overlap of $0\%$.
With the four-band rhythm information from all available channels, the recognition results of the DEAP and DREAMER are illustrated in Table \ref{table_win}.
With 1s temporal size of window with $25\%$ overlap, we have accuracies/standard deviations(\%) of 99.37/0.73, 99.35/0.91, 99.96/0.07, 99.93/0.07, and 99.95/0.07 for the five recognition tasks based on two datasets.
While in 2s with $25\%$ overlap case, we have accuracies/standard deviations(\%) of 95.17/0.45, 94.52/0.73, 99.53/0.06, 99.41/0.08, and 99.65/0.08.
The 3s we use $0\%$ overlap we have accuracies/standard deviations(\%) 89.03/4.83, 89.04/4.53, 98.70/0.20, 97.35/0.30, and 98.83/0.18.
The 4s we use $25\%$ overlap we have accuracies/standard deviations(\%) 74.14/0.60, 75.56/0.72, 98.32/0.19, 97.16/0.25, and 98.87/0.30.
As shown in Table \ref{table_win}, we achieve the highest recognition accuracy with a $1s$ temporal window length in DEAP-A, DEAP-V, DREAMER-A, DREAMER-V, and DREAMER-D, which are better than other longer temporal window lengths.
The performance reduced when the temporal window turns too long is due to the increasing of EEG signals complexity as the temporal size increases, which holds the same conclusion as in the previous studies.

In addition, we consider the 0.5s case to check the capability in tracking the small changes, with embedding parameters of $d=3, \tau=5$ (the 0.5s-window segments contain only 64 points), and overlapping equals to 0. We accomplish accuracies/standard deviations(\%) of 97.93/0.03, 98.28/0.03, 99.93/0.09, 99.92/0.10, and 99.82/0.53 for the five tasks.
The detailed results of DEAP subjects are illustrated in the supplement file of Table A.IV (0.5s),  Table A.I (1s), Table A.V (2s), Table A.VI (3s), and Table A.IV (4s).
The DREAMER subjects' results are illustrated in Table A.VIII(0.5s), Table A.II (1s), Table A.III (continue of 1s), Table A.IX(2s), Table A.X (3s), and Table A.XI(4s) of the supplement file.

\begin{table}[]
\caption{Exp \#4: Recognition Results with Single Channel EEG in DEAP Dataset.}
\begin{center}
\begin{tabular}{ccccccccc}
\toprule
  Number & Channel Content	& LA/HA(\%) & LV/HV(\%)\\
\midrule
1& Fp1& 	90.60/4.55 & 89.93/4.28		     \\
2&AF3 & 90.94/4.18	& 90.24/4.78    \\
3&F3 & 91.36/4.30	& 90.34/4.76    \\
4&F7 & 90.94/3.42	& 90.49/3.28    \\
5&FC5 & 91.14/3.90	& 89.98/4.60    \\
6&FC1 & 90.39/4.52	& 89.84/4.82    \\
7&C3 & 90.26/4.35	& 88.92/5.36    \\
8&T7 & 89.92/4.15	& 89.28/4.50    \\
9&CP5 &90.70/3.41& 89.55/4.67    \\
10&CP1 & 89.73/4.10	& 89.29/3.90    \\
11&P3 & 91.21/3.46	& 90.46/3.83   \\
12&P7 & 90.15/3.15	& 89.56/4.53  \\
13&PO3 & 90.65/4.37	&89.79/4.62  \\
14&O1 & 91.18/3.41	 & 90.74/4.22  \\
15&Oz & 91.74/4.41 & 90.73/4.58   \\
16&Pz & 91.09/3.17 & 90.68/3.66   \\
17&Fp2 & 89.62/4.16& 88.65/4.43    \\
18&AF4 &90.32/4.72 & 89.31/4.55 \\
19&Fz & 90.34/3.79	& 89.34/4.55   \\
20&F4 & 90.53/4.43	& 89.81/5.05    \\
21&F8 & 90.58/3.80	& 90.02/4.43    \\
22&FC6 & 90.82/3.92	& 89.32/5.59  \\
23&FC2&90.39/4.17 & 89.89/4.48\\
24&Cz & 	90.39/3.71& 89.68/4.27   \\
25&C4 & 91.39/3.99	& 90.49/4.28    \\
26&T8 &  90.47/4.37	& 89.09/5.10  \\
27&CP6 & 90.49/4.41  &89.76/4.75\\
28&CP2 & 90.02/3.40	& 88.90/4.43    \\
29&P4 &90.22/4.51	& 89.13/4.30   \\
30&P8 & 90.77/3.63	& 90.32/3.73    \\
31&PO4 & 89.74/4.17	& 89.10/4.95    \\
32&O2 & 91.12/4.09	& 90.33/3.48  \\
\midrule
&Mean&90.60/0.52&89.78/0.59\\
\bottomrule
\end{tabular}
\end{center}
\label{tab_sc_deap}
\end{table}

\begin{table}[]
\caption{Exp \#4: Recognition Results with Single Channel EEG in DREAMER Dataset.}
\begin{center}
\begin{tabular}{llccc}
\toprule
 Number& Channel 	& LA/HA (\%) &LV/HV (\%)  &LD/HD (\%)    \\
\midrule
1 & AF3	& 98.36/0.92 &98.23/0.96   & 98.19/0.92 \\
2 & F7	& 98.42/0.95  & 		98.33/1.01					&98.35/0.95  \\
3 & F3	& 98.95/0.44  &98.64/0.74&98.76/0.44  \\
4 & FC5& 98.65/0.89  &	98.38/1.01				 	&98.43/0.89  \\
5& T7& 98.66/0.89	&  			98.80/0.64				&98.72/0.71   \\
6 &  P7& 98.79/0.79 &  		98.78/0.83					& 98.82/0.79\\
7 &  O1& 97.73/1.68 &  		97.56/1.55		& 97.57/1.68  \\
8 &  O2& 98.96/0.81 & 		98.69/0.94		 & 98.80/0.81\\
9& P8 & 98.80/0.92	&			98.91/1.01		&  98.96/0.92   \\
10 & T8& 98.13/2.99 &  			98.08/3.18	&98.23/2.99  \\
11 &FC6& 98.59/1.09 &  		98.63/0.92	&98.80/1.09 \\
12&F4& 98.73/0.75	&			98.79/0.73		&  98.62/0.75   \\
13 & F8& 98.37/1.11 &  			98.84/0.88	& 98.39/1.11  \\
14 &  AF4	& 98.02/1.26 & 		97.88/1.18	 & 97.93/1.26  \\
\midrule
&Mean&98.51/0.36&98.44/0.39&98.47/0.39\\
\bottomrule
\end{tabular}
\end{center}
\label{tab_sc_dreamer}
\end{table}

\subsection{Exp. \#4: Emotion Recognition Comparison Using Single Channel EEG}
Most emotion-recognition BCI systems use multiple channels for feature extraction of dynamical functional connectivity analysis.
Though the multi-channel settings contain much more information than the single-bipolar EEG channel case, the burden brought by the electrodes restricts the application in wearable systems for lightweight applications.
Single-bipolar EEG channel settings can significantly reduce the complexity of emotion-recognition-based BCI systems. An example can be refereed in \cite{taran2019emotion}.
This work considered the single-channel EEG-based emotion recognition task with two datasets using ground reference electrode points in the 10-20 system.

In Exp. \#4, we systematically study the TEEGNDA analysis's channel variations, including emotion recognition experiments with single channels. 
The features involved in Exp \#4 are based on the combination of four rhythm bands. 
The results proposed are based on the average accuracies of 32 subjects from the DEAP dataset, while 23 subjects from the DREAMER dataset.
As Table \ref{tab_sc_deap} illustrates, the single channel-based DEAP-A, DEAP-V tasks perform worse than the combination-of-all situation.
However, the average accuracy(\%) with single-channel EEG information is 90.60 with a standard deviation(\%) of 0.52 in LA/HA-DEAP task and 89.78/0.59(\%) in single channel-based LV/HV-DEAP task.
The single-channel experiments results for DREAMER are illustrated in Table \ref{tab_sc_dreamer}, with average accuracies/standard deviations(\%) of 98.51/0.36, 98.44/0.39, and 98.47/0.39 for DREAMER-A, DREAMER-V, and DREAMER-D, respectively, lower than the multiple-channel fusion case.

\subsection{Exp. \#5: Emotion Recognition with Multi-Class Assessments}
In Exp. \#5, we consider two multi-class emotion recognition tasks based on the valence, arousal, and dominance levels.
Each emotion coordinate's high/low level in the valence-arousal-dominance model could be mapped into the Plutchik Wheel emotion model \cite{fourati2020unsupervised} as mentioned above.
Here we consider such two tasks from the involved dataset: the 4-class classification in DEAP and the 8-class classification in DREAMER.
The TEEGNDA framework built with RF classifier based on $1$-s sliding window size of 25\% overlap is performed on DEAP and DREAMER for the multi-class classification tasks. The PSR parameters are the as previous $1s$ case with $d = 8$ and $\tau = 10$.

As illustrated in Table \ref{table_mult_class}, we use the average values of overall accuracies, mean precisions, mean recalls, and mean F1-Scores of the multiple emotion classes as the assessments of the model.
For the 4-class classification task to distinguish LALV, LAHV, HALV, and HAHV labels in the DEAP dataset, the average recognition accuracy of the 32 subjects is 99.00\% with a standard deviation of 1.38\%. The average values/standard deviations of 4-class recall and F1-Scores are 98.36/2.30\% and 98.80/1.67\%. 
For the 8-class classification tasks in DREAMER, the average recognition accuracy of the 23 subjects is 99.89\% with a standard deviation of 0.20\%. The average values/standard deviations of 8-class recall and F1-Scores are 98.86/0.31\% and 99.89/0.22\%. 
The illustrated results proved that the proposed approach still shows good discrimination ability in multiple emotional state recognition with the topological features.

\begin{table}[]
\caption{Exp \#5: Classification Evaluations Performed with 1s Window (25\% Overlap) based on RF Classifier in 4-Class DEAP and 8-Class DREAMER.}
\begin{center}
\begin{tabular}{llccccccc}
\toprule
Tasks &  Accuracy(\%) & Precision(\%) & Recall(\%) & F1-Score(\%)  \\
\midrule
4-Class &  99.00/1.38 & 99.32/0.95 & 98.36/2.30 & 98.80/1.67\\
8-Class &   99.89/0.20 & 99.93/0.12 &99.86/0.31 & 99.89/0.22 \\
\bottomrule
\end{tabular}
\end{center}
\label{table_mult_class}
\end{table}

\section{Discussion}\label{sec:discussion}
EEG-based emotional state recognition contributes remarkably to a better understanding of human affections.
Exploring the nonlinear dynamical system-based EEG features has been previously investigated using descriptors such as entropy, geometrical parameters, fractal dimensions.
This work explores the topological properties of the nonlinear phase spaces of EEG sub rhythm band signals.
We found that the topological features extracted show excellent distinguishing ability in EEG-based emotional state recognition with the persistent homology technique.
Moreover, the EEG-based emotion recognition comparative studies of rhythm band, window size, and channel are also performed for comparisons, illustrating the proposed approach's robustness.
The proposed topological nonlinear dynamics analysis scheme provides an alternative descriptor compared to standard widely adopted features such as differential entropy (DE), power spectral density (PSD), asymmetry (ASM), differential asymmetry (DASM), differential caudality (DCAU).
Meanwhile, the TEEGNDA approach also shows competitive recognition ability compared to other recently proposed techniques.
In this section, we first compare our results with some of the previous related studies in EEG-based emotion recognition, and then we illustrate the technique details, while finally, we discuss method limitations and potential future directions.

\begin{table*}[]
\caption{Related Work with Frequency Rhythm Band-based Information}
\begin{center}
\begin{tabular}{llccccc}
\toprule
Dataset &  Feature Set or Framework & LA/HA(\%) & LV/HV(\%) &  LD/HD(\%)  & Window-Size  \\
\midrule
DEAP &DE Features + SparseDGCNN \cite{zhang2021sparsedgcnn} &   91.75/5.23 &95.72/3.75 & -  &  2s  \\
      &PSD Features + SparseDGCNN \cite{zhang2021sparsedgcnn} &\textbf{98.16/3.10}  &\textbf{98.74/1.61}  & - &  2s \\
      &DASM Features + SparseDGCNN \cite{zhang2021sparsedgcnn}  & 85.90/6.22 & 90.08/7.08 &  -&  2s\\
      &RASM Features + SparseDGCNN \cite{zhang2021sparsedgcnn}  & 85.50/8.35 & 90.08/9.77 & -& 2s\\
      &ASM Features + SparseDGCNN \cite{zhang2021sparsedgcnn}  & 95.17/5.80 & 90.89/6.40 & -& 2s\\
      &DCAU Features + SparseDGCNN \cite{zhang2021sparsedgcnn}  & 86.94/6.54 & 88.39/6.40 &-  &  2s\\
      &RACNN  \cite{cui2020eeg} & 97.11/2.01& 96.65/2.65 &  -& 1s\\ 
      &PCRNN  \cite{prcnn} & 91.03/2.99 & 90.80/3.08 &  -& 1s\\ 
      &GCNN  \cite{tao2020eeg} & 87.72/3.32& 88.24/3.18 &  -& 3s\\
      &CNN-RNN \cite{tao2020eeg} & 67.12/9.13 &62.75/7.53 &  -&3s \\
      &CNN-RNN-A  \cite{tao2020eeg} & 89.96/5.93 &  89.15/6.66& - & 3s\\
      &ACRNN  \cite{tao2020eeg}& \textbf{93.38/3.73} & \textbf{93.72/3.21} &  - &3s \\
       & \textbf{TEEGNDA}  & 89.03/4.83&  89.04/4.53& - &  3s   \\
      & \textbf{TEEGNDA}  & 95.17/0.45 & 94.52/0.73& - &  2s   \\
      & \textbf{TEEGNDA}  & \textbf{99.37/0.73} & \textbf{99.95/0.07} & - &  1s   \\
\midrule
DREAMER &DE Features + SparseDGCNN \cite{zhang2021sparsedgcnn} & 92.75/5.23 	& 96.64/3.32  & - & 2s \\
           &PSD Features + SparseDGCNN \cite{zhang2021sparsedgcnn}  & 98.55/2.72 & 99.18/1.20 & - & 2s\\
      &DASM Features + SparseDGCNN \cite{zhang2021sparsedgcnn} & 88.28/6.15& 92.34/6.70 &-&  2s \\
      &RASM Features + SparseDGCNN \cite{zhang2021sparsedgcnn}  &  86.88/8.32 & 91.27/9.45 & -&  2s\\
      &ASM Features + SparseDGCNN \cite{zhang2021sparsedgcnn}  & 95.84/5.30 & 92.62/8.63 & -&  2s\\
      &DCAU Features + SparseDGCNN \cite{zhang2021sparsedgcnn} & 89.11/6.41 & 93.18/5.88 & -&  2s \\
      &PSD Features + DGCNN \cite{song2018eeg} & 84.54/10.18 & 86.23/12.29 & - &2s \\
      &RACNN \cite{cui2020eeg} & 84.54/10.18 & 86.23/12.29 & 85.02/10.25 &  3s\\
      &GCNN  \cite{tao2020eeg} & 88.79/3.86& 88.87/3.58 & 88.54/3.89 &  3s\\
      &CNN-RNN \cite{tao2020eeg} & 77.66/13.34& 78.59/13.87& 77.75/14.22&  3s\\
      &CNN-RNN-A  \cite{tao2020eeg} & 97.36/2.63& 96.61/3.42&97.54/2.16 &  3s\\
      &ACRNN  \cite{tao2020eeg}& 97.98/1.92&\textbf{97.93/1.73} & 98.23/1.42& 3s\\
       &\textbf{TEEGNDA}  & \textbf{98.70/0.20} & 97.35/0.30 & \textbf{98.83/0.18} &  3s  \\
       &\textbf{TEEGNDA}  & \textbf{99.53/0.06} & \textbf{99.41/0.08} & \textbf{99.65/0.08} &  2s  \\
\bottomrule
\end{tabular}
\end{center}
\begin{tablenotes}
\RaggedRight
\item[a]  
\end{tablenotes}
\label{table_compare}
\end{table*}

\subsection{Comparison with Related Work}
There are a variety of works proposed for emotional state classification, typical rhythm sub-band EEG based features are DE, PSD, ASM, DASM, and DCAU \cite{zheng2017identifying, zheng2015investigating, zhang2021sparsedgcnn}.
Recently, Zhang et al. \cite{zhang2021sparsedgcnn} proposed a sparse dynamic graph convolutional neural network (sparse DGCNN) framework to investigate the rhythm sub-band EEG-based emotion recognition. Comprehensive comparisons have been proposed using the DE, PSD, ASM, DASM, and DCAU features.
In \cite{song2018eeg}, Song et al. proposed a PSD features-based DGCNN framework using EEG rhythm band signals from DREAMER.

Despite the EEG rhythm band-based frameworks, there are a variety of approaches have been developed based on the preprocessed raw EEG.
The recent fast-developing techniques of deep learning models adopted the representative ability of large-scale neural networks to reveal the nonlinearities of neural systems based on EEG.
Typical works involved DEAP and DREAMER are \cite{tao2020eeg, cui2020eeg, prcnn}, we achieve comparable results as well.
As Table \ref{table_compare} illustrated, with similar experimental settings, we achieve comparable results in the DEAP dataset while better results in the DREAMER dataset.
The comparisons illustrate that a topological approach is a powerful tool in understanding the nonlinear dynamics of the neural systems toward EEG-based emotion recognition.
However, it is not possible to cover all the techniques of EEG-based emotional state recognition. We only illustrate recent typical works both in a sub-rhythm band EEG-based and raw EEG-based.
The comparisons validated the effectiveness of our TEEGNDA framework developed with sub rhythm band EEG signals.

\subsection{Comparison with Other Nonlinear Dynamics Descriptors}
The TEEGNDA approach is developed based on describing the nonlinear dynamics revealed in the phase space. We introduced some related works using other descriptions such as entropy-based and geometrical representation-based approaches in the introductions.
Compared to current widely used descriptors, the TDA technique provides a choice to extract information from the phase space. Namely, we term it as topological nonlinear dynamics analysis, which shows excellent representative ability to classify the emotional states.
In order to show the superiorities, we perform the rhythm band analysis-based emotion recognition tasks using six typical nonlinear descriptors: fuzzy entropy, approximate entropy, sample entropy, recurrent plot, Poincare plot, and Lyapunov exponents (with 1s sliding window length and 25\% overlap, implemented in the same way as in the TEEGNDA framework, by replacing the rhythm band signal feature extracting with these six nonlinear parameter calculation).
The parameter of each approach is set as the same in our approach to guarantee fairness in the comparisons.
As presented in Table \ref{table_compare_nd}, the TDA technique outperforms the other nonlinear descriptors, including entropy-based and geometry-based ones.

\begin{table*}[]
\caption{Comparison with Current EEG Nonlinear Dynamics Descriptors}
\begin{center}
\begin{tabular}{llcccccc}
\toprule
Dataset &  Nonlinear Dynamics Descriptor & LA/HA(\%) & LV/HV(\%) &  LD/HD(\%)  & 4-Class Emotion(\%) & 8-Class Emotion(\%) \\
\midrule
DEAP &  Fuzzy Entropy &   69.26/5.96 &71.57/6.91& -  &  54.20/6.80& -  \\
      &  Approximate Entropy&68.77/6.52 & 70.64/6.71 &  - & 52.68/6.84 & - \\
      & Sample Entropy & 67.95/6.05 & 69.36/7.11 &  -&  52.43/6.39& - \\
      & Recurrence Plot &58.39//6.24  & 64.03/9.64  & -& 41.02/6.37 & - \\
      & Poincare Plot & 67.97/18.78&69.51/19.27& -& 55.54/16.05& - \\
       & Lyapunov Exponent & 57.58/7.05& 63.67/10.96& -& 39.24/7.75& - \\
      & \textbf{TEEGNDA}  & \textbf{99.35/0.74} & \textbf{99.33/0.76} & - & \textbf{99.00/1.38}  & -  \\
\midrule
DREAMER & Fuzzy Entropy &   81.90/6.50 &81.89/6.48 & 81.25/6.95  &  -&69.77/1.04\\
      & Approximate Entropy & 81.40/6.36 & 81.19/6.28  & 80.47/6.56 &  - & 68.47/9.60 \\
      & Sample Entropy & 80.37/6.13 & 79.42/6.75 &79.41/6.82 &  - & 66.10/10.03 \\
      & Recurrence Plot & 71.63/8.72 & 70.27/9.17 & 70.22/9.65 & - & 48.97/14.47\\
      & Poincare Plot & 87.17/6.52 & 86.39/7.14 & 86.92/6.66 & - & 78.68/10.46 \\
       & Lyapunov Exponent &  64.77/9.32 & 61.89/8.21 & 64.26/8.36 & - & 37.88/11.42 \\
        & \textbf{TEEGNDA} & \textbf{99.92/0.12} & \textbf{99.92/0.12} & \textbf{99.95/0.08} &- &  \textbf{99.89/0.20}  \\
\bottomrule
\end{tabular}
\end{center}
\begin{tablenotes}
\RaggedRight
\item[a]  
\end{tablenotes}
\label{table_compare_nd}
\end{table*}

\subsection{Model Parameters Discussion}
The EEG signals of different frequency bands were embedded into the phase space via time-delay embedding with fixed time-delay $\tau$ and dimension $d$.
The embedding parameters determined how well the nonlinear dynamics revealed in the phase space could potentially impact the nonlinear models developed with statistical parameters or geometrical representation.
The changing of embedding parameters causes the geometrical structure variation in phase space. 
The topological descriptors consider the state points' connection relationships underlaying the phase space, which are more robust than the geometrical descriptors such as recurrent or Poincare plots. 
The tracked topological objects, such as the 1-dimensional homologies (holes) in the filtration process, are less impacted by the embedding parameters.

\subsection{Limitations and Potential Improvement}
The subject-independent analysis has been widely considered in the EEG-based emotion recognition works, which were used to show the recognition ability of general patterns in EEG.
Significantly, modern deep learning techniques provide a powerful tool to represent such kinds of cross-subject features as in \cite{zheng2015investigating, zhang2021sparsedgcnn, zhong2020eeg} via the network structure.
In the current work, the individual difference in EEG dramatically impacts the final recognition accuracy.
Firstly, the emotional rating scores recorded through watching emotional film clips are different in different subjects in the emotion score site. The valence and arousal dimensions are different even using the same stimuli, caused by the differences in inner psychological characteristics. 
Secondly, in the physiological characteristics site, as a sensitive and real-time physiological model, EEG signals could vary from individual to individual due to their unique internal physiological characteristics \cite{zhang2021sparsedgcnn}.
Thirdly in the nonlinear modeling site, the proposed model explores the topological description of the point clouds in the frequency-band EEG phase space, which reveals the nonlinear dynamics of the brain neural system, which is supposed to be mainly impacted by the individual physiological characteristics.
The topological descriptions give subtle structure information of the point cloud in the phase space, as each point of the cloud illustrates a potential state of the nonlinear system revealed the personality and individual differences. 
Thus, the subject-independent evaluation cannot deal with the three individual differences, which is supposed to be the main limitation in the subject-independent analysis compared to other feature-fusion systems and deep learning structures.




\section{Conclusion}\label{sec:conclusion}
In this paper, we proposed a topological nonlinear dynamics analysis scheme for EEG-based emotion recognition together with the TEEGNDA framework.
The EEG signals reveal the brain dynamics as measurements of a nonlinear system in the dynamical system theory.
The emotion states represented with arousal, valence, and dominance level are distinguished by investigating the phase spaces' topological properties of different EEG rhythm bands.
The proposed TEEGNDA approach adopts persistent homology techniques to extract topological features from different EEG rhythm bands to build feature vectors toward emotion classification tasks.
The results demonstrate that the TEEGNDA approach performs excellently in the subject-wise experiments in the DEAP and DREAMER datasets, namely with average accuracy/standard deviation (\%) of 99.37/0.73 for DEAP-A, 99.35/0.91 for DEAP-V, 99.96/0.07 for DREAMER-A, 99.93/0.07 for DREAMER-V, and 99.95/0.07 for DREAMER-D, which show great performance and competitive to other models using the similar experimental settings.
Furthermore, we compared the performance of the approach using different EEG rhythms bands, temporal windows, and single-channel choices.
The proposed approach also shows good recognition ability with single-channel EEG signals.
The topological features bring an alternative tool toward EEG signal analysis and brain dynamics analysis.

\ifCLASSOPTIONcaptionsoff
  \newpage
\fi



%

%


\end{document}